\documentclass[preprint,aps]{revtex4-1}
\usepackage{hyperref}
\usepackage{graphicx}
\usepackage{epstopdf}
\usepackage{subfig}
\usepackage{amsmath}
\usepackage{appendix}

\def\be{\begin{equation}}
\def\ee{\end{equation}}
\def\bea{\begin{eqnarray}}
\def\eea{\end{eqnarray}}
\def\ba{\begin{array}}
\def\ea{\end{array}}
\def\bc{\begin{center}}
\def\ec{\end{center}}
\begin{document}
\title{Charge and magnetization densities in transverse coordinate and impact parameter space}
\author{Narinder Kumar and Harleen Dahiya}
\address{Department of Physics\\
Dr. B. R. Ambedkar National Institute of Technology\\
         Jalandhar-144011, India}
\begin{abstract}
Electromagnetic form factors obtained from the overlap of light front wave functions (LFWFs) have been used to study the transverse densities of charge and magnetization. The calculations have been carried out to develop a relation between the charge distribution of the quarks inside nucleon in the transverse coordinate space as well as in the impact parameter space. When a comparison is carried out, it is found that the transverse distribution in the impact parameter space, where the longitudinal momentum fraction $x$ can be fixed, falls off faster than
the spatial distribution in the transverse coordinate space where there is some contribution from the longitudinal momentum as well. The anomalous magnetization density of the nucleon has also been discussed. Further, we have also presented the results of the QCD transverse AdS charge density inspired from the holographic QCD model.
\end{abstract}
\maketitle
\section{introduction}
Study of form factors has been important in the field of hadron physics as they provide important information about internal structure of the nucleon \cite{jeffer,jeffer1,babar,sample,a4,happex,emff}. Recently, experiments have measured and produced remarkably precise data for the electromagnetic form factors \cite{emff1,emff2,emff3,emff4}. The nucleon electromagnetic form factors have been used to study the distribution of charge and magnetization densities of quarks inside the nucleon \cite{gamiller,dshwang,ovteryeav}. To get a more clear picture of the nucleon structure, one can fully parameterize the structure of the nucleon in terms of generalized parton distributions (GPDs) \cite{diehl,ji,radyushkin} which are functions of longitudinal momentum $x$ of the quark, the invariant momentum transfer $t$ and the skewness parameter $\xi$ which gives the fraction of the longitudinal momentum transfer to the nucleon. The Fourier transform of the GPDs w.r.t transverse momentum transfer has not only been used to study the nucleon structure in transverse impact parameter space \cite{impact,impact1,impact2,impact3,impact4} but it also gives a tomographic picture of the distribution of quark charge densities inside the nucleon through the impact parameter dependent parton distribution function (ipdpdf) $q(x,b_\perp)$ for a quark of momentum fraction $x$ located at a transverse position $b_\perp$. It has been shown that ipdpdf of the transversely localized states of the same helicity is the two-dimensional Fourier transform of the spin non-flip GPD $H$ \cite{impact1,impact3}.  The role of transverse momentum of parton has been significant to understand the nucleon structure as it yields information on the fragmentation functions and the single spin asymmetries (SSAs) \cite{fsi,fsi1}.  Some efforts have been made to understand this phenomena by taking one gluon exchange in the final state interactions when the distribution functions depend on the longitudinal as well as transverse momenta of the partons \cite{distort,distort1,distort2}.

Transverse charge density defines a way to analyze electromagnetic form factors of hadrons moving at infinite momentum with transverse distance $b_\perp$ from the transverse center of momentum \cite{soper}.  Charge densities have been discussed in impact parameter space for the polarized and transversely polarized nucleons \cite{gamiller,ovteryeav,negachargedense,chargedensity_p_n,chargedensity_Ndelta_transition},  in finite radius approximation \cite{venkat}, in scalar diquark model using transverse coordinate space \cite{dshwang} as well as using Sachs form factors \cite{sachs,kelly}. Transverse densities can also be useful in the study of the spatial distribution of the momentum $P^+$ and can be related with the Fourier transforms of the gravitational form factors $A(q^2)$ and $B(q^2)$ \cite{carlson}. In addition to this, magnetization density can be evaluated to get information on the anomalous magnetic moment of the nucleon \cite{magmoment}. Recently, transverse charge and magnetization densities have also been studied in the holographic QCD \cite{holographicqcd,adsspace}.

The above discussion motivate us to study the charge and magnetization densities obtained from the helicity non flip and helicity flip form factors. One of the most successful model which finds its application for the quantities discussed above consists of spin-$\frac{1}{2}$ system as a composite of spin-$\frac{1}{2}$ fermion and spin-1 vector boson.  We have generalized the framework of QED by assigning a mass $M$ to external electrons in the Compton scattering process, but a different mass $m$ to the internal electron line and a mass $\lambda$ to the internal photon line. The idea behind this is to model the structure of a composite fermion state with a mass $M$ by a fermion and a vector constituent with respective masses $m$ and $\lambda$ \cite{brodsky,brodsky1,dipankar}. In our case we take $\xi=0$ \cite{impact1,impact4} which represents the momentum transfer exclusively in transverse direction leading to the study of ipdpdfs in transverse impact parameter space. This model serve as a template to study the internal structure of nucleon and is self consistent since it has correct correlation of different Fock components of the state as given by light front eigen value equation. This model has been used to calculate the spin and orbital angular momentum of a composite relativistic system as well as the GPDs in impact parameter space and it also gives the Schwinger anomalous magnetic moment and the corresponding Dirac's and Pauli form factor including the vanishing of the anomalous gravitomagnetic moment.

The purpose of the present communication is to determine the distribution of nucleon charge density in transverse coordinate space and in impact parameter space using light front wave functions (LFWFs) of simple spin-$\frac{1}{2}$ objects like dressed quark or a dressed electron in theory. We begin by computing the electromagnetic Dirac and Pauli form factors as well as the Sachs form factors $(G_E(Q^2))$ and $(G_M(Q^2))$ from the overlap of LFWFs. The anomalous magnetization density of the nucleon obtained from the spin flip Pauli form factor has also been discussed.
Further, it would be significant to develop a relation between the  distribution of  quark charge densities inside nucleon in the transverse coordinate space as well as in the impact parameter space particularly for the $u$ and $d$ quark distribution by using a fitted set of parameters. This will give a deeper understanding of both the spatial and transverse distribution of the quark densities.
Furthermore, it would be interesting to extend the calculations to present the predict the results of the QCD transverse AdS charge density inspired from the holographic QCD model.

\section{Nucleon charge densities}
The nucleon form factors are defined in terms of the electromagnetic current operator $J^\mu(x)$ as \cite{diehl,vander-em,arrington-em,punjabi-em}
\be
\langle p', \lambda'| J^\mu(0)|p,\lambda \rangle= \bar{u}(p',\lambda')\left(\gamma^\mu F_1(q^2) +i \frac{\sigma^{\mu \nu}}{2 M} q_\alpha F_2(q^2)\right)u(p,\lambda),
\ee
where $q_\nu=p_\nu'- p_\nu$ is the momentum transfer and in the present case it is taken as space-like ($q^+=0$) leading to $-q^2=Q^2$.  The $\sigma^{\mu \nu}$ is the spin operator, $\gamma^\mu$ denote the Dirac matrices and $u(p,\lambda)$ for the Dirac spinor with mass $M$. The initial (final) momentum and helicity of the nucleon are taken as $p,\lambda$ $(p',\lambda')$  respectively. The Dirac and Pauli form factors ($F_1(q^2)$ and $F_2(q^2)$) are normalized so that $F_1(0)$ and $F_2(0)$ are the nucleon charge and the anomalous magnetic moment at zero momentum transfer respectively. The Sachs form factors \cite{sachs} can be defined in terms of the Dirac and Pauli form factors as follows
\be
G_E(q^2)=F_1(q^2)-\frac{q^2}{4 M^2}F_2(q^2),
\label{GE}
\ee
and
\be
G_M(q^2)=F_1(q^2)+F_2(q^2).
\label{GM}
\ee
Transverse charge density for a unpolarized nucleon $\rho(b_\perp)$ gives the charge density of the nucleon at a transverse position $b_\perp$ for any value of the longitudinal momentum or position. It can can be obtained by integrating the ipdpdfs over the longitudinal momentum fraction $x$ and can be expressed in terms of the Dirac form factor \cite{gamiller,dshwang} as follows
\be
\rho(b_\perp)=\int \frac{d^2 q}{(2 \pi)^2} F_1(q^2) e^{-i \vec{q}_\perp \cdot \vec{b}_\perp}.
\label{rho_density}
\ee
In terms of $G_E$ and $G_M$ the transverse charge density for unpolarized nucleon  can be expressed as
\be
\rho(b_\perp)=\int_{0}^{\infty}  \frac{q \ dq}{2 \pi} J_0(q b_\perp) \frac{G_E(q^2)+\tau G_M(q^2)}{1+\tau},
\label{tran_charge_density}
\ee
where $J_0$ is a cylindrical Bessel function and $\tau=\frac{q^2}{4 M^2}$.
The main advantage of charge density in impact parameter space is that it gives the density interpretation when one compares it with the Fourier transform of the Sachs form factor. The density in impact parameter space is calculated from the Dirac form factor which is further obtained by integrating the GPDs over the longitudinal momentum fraction $x$. The GPDs have been studied \cite{impact} in impact parameter space where the parton distributions are expressed in terms of impact parameter dependent wave functions using Lorentz invariance to describe the transverse structure of a fast moving hadron. These distributions are the matrix elements of the quark operators between nucleon states with different  momenta. The GPDs are of high interest as they can be related to the total angular momentum carried by the quarks in the nucleon and can be directly found from the Deep Virtual Compton scattering (DVCS) experiments \cite{dvcs}.

The Dirac and Pauli form factors can be expressed as overlaps of the light front wave functions (LFWFs) for the two-particle Fock state.
We consider here a spin-$\frac{1}{2}$ system as a composite of spin-$\frac{1}{2}$ fermion and spin-1 vector boson. The details of the model have been presented in Ref. \cite{brodsky,brodsky1,dipankar}, however, for the sake of completeness we present here the essential two-particle wave functions for spin up and spin down electron  expressed as
\begin{eqnarray}
&&\psi_{+\frac{1}{2}+1}^{\uparrow}(x,\vec{k}_\perp)=-\sqrt{2}\frac{-k^1+\iota k^2}{x(1-x)}\varphi,\nonumber\\ && \psi_{+\frac{1}{2}-1}^{\uparrow}(x,\vec{k}_\perp)=-\sqrt{2}\frac{k^1+\iota k^2}{(1-x)}\varphi,\nonumber\\ &&
\psi_{-\frac{1}{2}+1}^{\uparrow}(x,\vec{k}_\perp)=-\sqrt{2}\Big(M-\frac{m}{x}\Big)\varphi,\nonumber\\ &&\psi_{-\frac{1}{2}-1}^{\uparrow}(x,\vec{k}_\perp)=0 \,,
\label{spinup}
\end{eqnarray}
and
\begin{eqnarray}
&&\psi_{+\frac{1}{2}+1}^{\downarrow}(x,\vec{k}_\perp)=0,\nonumber\\&&
\psi_{+\frac{1}{2}-1}^{\downarrow}(x,\vec{k}_\perp)=-\sqrt{2}\Big(M-\frac{m}{x}\Big)\varphi,\nonumber\\ &&\psi_{-\frac{1}{2}+1}^{\downarrow}(x,\vec{k}_\perp)=-\sqrt{2}\frac{-k^1+\iota k^2}{(1-x)}\varphi,\nonumber\\
&&\psi_{-\frac{1}{2}-1}^{\downarrow}(x,\vec{k}_\perp)=-\sqrt{2}\frac{k^1+\iota k^2}{x(1-x)}\varphi \,,
\label{spindown}
\end{eqnarray}
with the energy denominator defined as
\begin{eqnarray}
\varphi(x, \vec{k_{\perp}}) =\frac{e }{\sqrt {1-x}} \frac{1}{\left(M^2-\frac{\vec{k}_{\perp}^{2}+m^2}{x}-\frac{\vec{k}_{\perp}^{2}+\lambda^2}{1-x}\right)}\,.
\end{eqnarray}
Here the formalism has been generalized by assigning a mass $M$ to external fermions in the Compton scattering process but a different mass $m$ to the internal fermion line and a mass $\lambda$ to the internal boson line.
The energy denominator can be further generalized by adjusting its power behaviour {\it p} \cite{dshwang,mulders} as follows
\begin{eqnarray}
\varphi(x, \vec{k_{\perp}}) =\frac{e M^{2p}}{x^p(\sqrt {1-x})} \left(M^2-\frac{\vec{k}_{\perp}^{2}+m^2}{x}-\frac{\vec{k}_{\perp}^{2}+\lambda^2}{1-x}\right)^{-p-1}\,,
\end{eqnarray}
the term $M^{2p}$ has been added for dimensionality and $x^p$ to satisfy the polynomiality condition. In the present work we have taken $p=1$.

\begin{figure}
    \includegraphics[width=6cm]{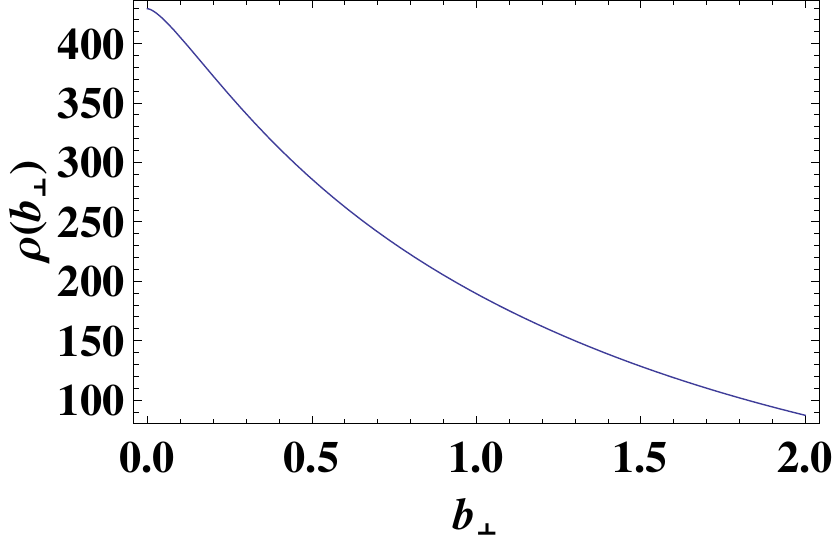}
\caption{Nucleon charge density (in $fm^{-2}$) as a function of $b_\perp$ (in $fm$).}
 \label{nucleon_density}
\end{figure}

The Dirac and Pauli form factor defined in terms of the spin non-flip and spin flip matrix elements of the $J^+$ current \cite{brodsky1} are expressed as follows
\begin{eqnarray}
F_1(q^2)&=&
\left<\Psi^{\uparrow}(P^+, {\vec P_\perp}= \vec q_\perp)
|\Psi^{\uparrow}(P^+, {\vec P_\perp}= \vec 0_\perp)\right>
\nonumber\\
&=& \int\frac{{\mathrm d}^2 {\vec k}_{\perp} {\mathrm d} x }{16 \pi^3}
\Big[\psi^{\uparrow\ *}_{+\frac{1}{2}\, +1}(x,{\vec k'}_{\perp})
\psi^{\uparrow}_{+\frac{1}{2}\, +1}(x,{\vec k}_{\perp})
+\psi^{\uparrow\ *}_{+\frac{1}{2}\, -1}(x,{\vec k'}_{\perp})
\psi^{\uparrow}_{+\frac{1}{2}\, -1}(x,{\vec k}_{\perp})
\nonumber\\
&&\qquad\qquad\qquad\qquad
+\psi^{\uparrow\ *}_{-\frac{1}{2}\, +1}(x,{\vec k'}_{\perp})
\psi^{\uparrow}_{-\frac{1}{2}\, +1}(x,{\vec k}_{\perp})
\Big]\ ,
\label{diracff}
\end{eqnarray}
\bea
F_2(q^2)&=& {-2M\over (q^1-{\mathrm i}q^2)} \left
<\Psi^{\uparrow}(P^+, {\vec P_\perp}= \vec q_\perp))
|\Psi^{\downarrow}(P^+, {\vec P_\perp}= \vec 0_\perp)\right>
\nonumber\\ &=&{-2M\over (q^1-{\mathrm i}q^2)} \int\frac{{\mathrm
d}^2 {\vec k}_{\perp} {\mathrm d} x }{16 \pi^3}
\Big[\psi^{\uparrow\ *}_{+\frac{1}{2}\, -1}(x,{\vec k'}_{\perp})
\psi^{\downarrow}_{+\frac{1}{2}\, -1}(x,{\vec k}_{\perp})
\nonumber\\
&& + \psi^{\uparrow\ *}_{-\frac{1}{2}\, +1}(x,{\vec k'}_{\perp})
\psi^{\downarrow}_{-\frac{1}{2}\, +1}(x,{\vec k}_{\perp}) \Big].
\label{pauliff}
\eea
The spin non-flip and spin flip GPDs can similarly be  expressed in terms of the spin up and spin down electron wave functions as
\bea
H(x,0,\vec{q}_\perp)&=& \int{\frac{d^2\vec{k}_\perp}{16\pi^3}} \bigg[ \psi_{+\frac{1}{2}+1}^{\uparrow *}(x,\vec{k}'_\perp) \psi_{+\frac{1}{2}+1}^{\uparrow}(x,\vec{k}_\perp)+\psi_{+\frac{1}{2}-1}^{\uparrow *}(x,\vec{k}'_\perp) \psi_{+\frac{1}{2}-1}^{\uparrow}(x,\vec{k}_\perp)+\nonumber\\
&& \psi_{-\frac{1}{2}+1}^{\uparrow *}(x,\vec{k}'_\perp) \psi_{-\frac{1}{2}+1}^{\uparrow}(x,\vec{k}_\perp) \bigg] \,,
\label{h2}
\eea
\be
\frac{\Delta^1- i \Delta^2}{2 M} E(x,0,\vec{q}_\perp)=\int{\frac{d^2\vec{k}_\perp}{16\pi^3}} \Big[\psi_{+\frac{1}{2}+1}^{\uparrow *}(x,\vec{k}'_\perp) \psi_{+\frac{1}{2}+1}^{\downarrow}(x,\vec{k}_\perp)+\psi_{+\frac{1}{2}-1}^{\uparrow *}(x,\vec{k}'_\perp) \psi_{+\frac{1}{2}-1}^{\downarrow}(x,\vec{k}_\perp) \Big].
\label{e2}
\ee
The Dirac and Pauli form factors obtained from Eqs. (\ref{spinup}), (\ref{spindown}), (\ref{diracff}) and (\ref{pauliff}) are given as
\bea
F_1(q^2)&=&\frac{e^2}{16 \pi^3} M^4 \int dx \Big(2 (1+x^2)(1-x) I_1 + \frac{2 (M x-m)^2 (1-x)^3}{x^2}I_2 \Big),
\eea
\bea
F_2(q^2)&=& - 4 M^5 \frac{e^2}{16 \pi^3} \int dx (M x-m)x(1-x)^3 I_2,
\eea
where
\bea
I_1&=& \pi \int_{0}^{1} \frac{\alpha (1-\alpha)}{2 D^2} d\alpha, \nonumber\\
I_2&=& \pi \int_{0}^{1} \frac{\alpha (1-\alpha)}{D^3} d\alpha, \nonumber\\
D&=& \alpha (1-\alpha) (1-x)^2 q_\perp^2 - M^2 x(1-x)+m^2 (1-x) + \lambda^2 x .
\label{integrals}
\eea
Using Eq. (\ref{GE}) and (\ref{GM}), we can also solve for the Sachs form factors and obtain
\begin{eqnarray}
G_E(q^2)&=&\frac{e^2}{16 \pi^3}M^4 \int dx  2 (1+x^2)(1-x) I_1 + \frac{2 (M x-m)^2 (1-x)^3}{x^2} I_2+ \nonumber\\
&& \frac{e^2}{16 \pi^3} Q^2 M^3 \int dx (M x-m) x (1-x)^3 I_2,
\end{eqnarray}
\bea
G_M(q^2)&=& \frac{e^2}{16 \pi^3} \Big(M^4 \int dx(2(1+x^2)(1-x)I_1+ \frac{2 (M x-m)^2 (1-x)^3}{x^2} I_2)- \nonumber\\
&& 4 M^5 \int (M x-m) x (1-x)^3 dx I_2 \Big).
\eea
The transverse charge density from Eq. (\ref{tran_charge_density}) in terms of Sachs form factors can now be written as
\bea
\rho(b_\perp)&=&\int_{0}^{\infty} \frac{dQ \ Q}{2 \pi} J_0(Q \ b_\perp) \int_{0}^{1} dx \frac{(1+\tau) \left( 2 M^4 (1+x^2) (1-x)I_1 +\frac{2 (M x-m)^2 (1-x)^3 I_2}{x^2}\right)}{1+\tau} + \nonumber\\
&& \frac{ Q^2 M^3 (M x-m)x (1-x)^3 I_2 - 4 \tau M^5 (M x-m) x (1-x)^3 I_2}{1+\tau}.
\eea
In Fig. \ref{nucleon_density} we have presented the result for nucleon charge density as a function of impact parameter.
The impact parameter space is phenomenologically important as it can be obtained from the GPDs ($H$ and $E$) which satisfy the Lorentz invariance principle. Even the constraint due to the invariance of the light cone formulation under transverse boost $b_\perp= \sum x_i b_{i \perp}$ identifies $b_\perp$ as transverse center of momentum of the partons in Fock state. In addition to this, impact parameter representation of the GPDs also satisfies the positivity constraints \cite{pobylista}. For numerical calculations we have taken the value of masses of external fermion, internal fermion and internal boson as $M=m=0.5$MeV and $\lambda=0.02$MeV respectively after imposing the  condition $M<m+\lambda$ to prevent nucleon decay \cite{mukherjee}. From the figure, we find that the charge density is peaked near $b_\perp=0$ and decreases as the value of $b_\perp$ increases which implies a maximum distribution near the center of momentum.
Similar technique has recently been used to extract the transverse pion and proton charge density from the pion form factor data \cite{carmingtoo} where it has been discussed that the data from JLab 12 GeV and EIC \cite{jlab1,jlab2} will increase the dynamic extent of the form factor data to higher values of $Q^2$ and thus reduce the uncertainties in the extracted pion transverse charge density.

\section{Charge density in transverse coordinate space}
The LFWFs in transverse coordinate space $\tilde{\psi}(x,\vec{r}_\perp)$ can be obtained by Fourier transforming $\psi(x,\vec{k}_\perp)$ in the momentum space
\bea
\tilde{\psi}(x,\vec{r}_\perp)&=&\int \frac{d^2 k_\perp}{(2 \pi)^2} e^{i \vec{k}_\perp \cdot \vec{r}_\perp} \psi(x,\vec{k}_\perp),
\label{fttrans}
\eea
where $\vec{r}_\perp$ is the transverse coordinate. The LFWFs for the spin up state in the transverse coordinate space can now be expressed as
\bea
\tilde{\psi}^\uparrow_{+\frac{1}{2}+1}(x,\vec{r}_\perp)
&=&  i \frac{e}{8 \pi} \sqrt{2} M^2 (1-x)^\frac{3}{2} r_\perp  K_0(\sqrt{D}  r_\perp), \nonumber\\
\tilde{\psi}^\uparrow_{+\frac{1}{2}-1}(x,\vec{r}_\perp)
&=& -i \frac{e}{8 \pi} \sqrt{2} M^2x \sqrt{1-x} r_\perp   K_0(\sqrt{D}  r_\perp), \nonumber\\
\tilde{\psi}^\uparrow_{-\frac{1}{2}+1}(x,\vec{r}_\perp)
&=& - \frac{e}{4 \pi} \sqrt{2} M^2 (M x-m)(1-x)^\frac{3}{2} \frac{1}{\sqrt{D}}r_\perp K_1(\sqrt{D} r_\perp), \nonumber\\
\tilde{\psi}^\uparrow_{-\frac{1}{2}-1}(x,\vec{r}_\perp)&=& 0,
\eea
whereas for spin down state we have
\bea
\tilde{\psi}^\downarrow_{+\frac{1}{2}+1}(x,\vec{r}_\perp)&=& 0, \nonumber\\
\tilde{\psi}^\downarrow_{+\frac{1}{2}-1}(x,\vec{r}_\perp)&=& - \sqrt{2} \frac{e}{4 \pi} M^2(M x-m)(1-x)^\frac{3}{2} \frac{r_\perp}{\sqrt{D}} K_1(\sqrt{D} r_\perp),\nonumber\\
\tilde{\psi}^\downarrow_{-\frac{1}{2}+1}(x,\vec{r}_\perp)&=& i \frac{e}{8 \pi} \sqrt{2} M^2 x (1-x)^\frac{1}{2} r_\perp K_0(\sqrt{D}  r_\perp) , \nonumber\\
\tilde{\psi}^\downarrow_{-\frac{1}{2}-1}(x,\vec{r}_\perp)&=& - i \frac{e}{8 \pi} \sqrt{2} M^2 r_\perp (1-x)^\frac{1}{2} K_0(\sqrt{D}  r_\perp),
\eea
where $K$ is the Bessel function of second kind.

Even though the nucleon charge density in impact parameter space ($\rho(b_\perp)$) is the more correctly defined quantity with proper invariance properties to be compared directly with the observables, we define the density distribution in the transverse coordinate space for the sake of comparison as follows
\bea
P(\vec{r}_\perp)&=& \int dx [ \tilde{\psi}^{\uparrow *}_{+\frac{1}{2}+1}(x,\vec{r}_\perp) \tilde{\psi}^\uparrow_{+\frac{1}{2}+1}(x,\vec{r}_\perp) + \tilde{\psi}^{\uparrow *}_{+\frac{1}{2}-1}(x,\vec{r}_\perp) \tilde{\psi}^\uparrow_{+\frac{1}{2}-1}(x,\vec{r}_\perp) + \nonumber\\
&& \tilde{\psi}^{\uparrow *}_{-\frac{1}{2}+1}(x,\vec{r}_\perp) \tilde{\psi}^\uparrow _{-\frac{1}{2}+1}(x,\vec{r}_\perp)],
\label{transcoor}
\eea
leading to
\bea
P(\vec{r}_\perp)&=& \int dx \frac{e^2}{16 \pi^2} \Big[\frac{1}{2} M^4 (1-x) r^2_\perp  (1+x^2) K_0^2(\sqrt{D} r_\perp) + 2 (M x-m)^2 M^4 \nonumber\\
&& \times \frac{(1-x)^3}{D} r^2_\perp K_1^2(\sqrt{D} r_\perp) \Big].
\eea

A precise relation between the nucleon density distribution $P(r_\perp)$ and charge distribution $\rho(b_\perp)$ can be obtained using elementary convolution theorems and Fourier transforms (details have been presented in  Appendix A) and is given as
\bea
\rho(x,\vec{b}_\perp)=  \frac{1}{(1-x)^2}P\left(x, \frac{\vec{b}_\perp}{-1+x} \right).
\label{relation}
\eea
In order to obtain the explicit contributions of the $u$ and $d$ quarks in the nucleon density distribution $P(r_\perp)$ and charge distribution $\rho(b_\perp)$, we have followed the formalism used in Ref. \cite{dshwang} where the isospin  symmetry $P_u=P_p+\frac{P_n}{2}$, $P_d= P_p + 2 P_n$ has been used with similar type of  relations for charge density in the impact parameter space. A fitting can be carried out with the experimental data of the Dirac form factors of the nucleon to obtain the fitted values of the masses of the internal fermions and bosons. For the present work we have used the same set of fitted values of $m$, $\lambda_u$ and $\lambda_d$ as in Ref.\cite{dshwang,kelly,kelly_sachs}.

\begin{figure}
 \minipage{0.42\textwidth}
  \includegraphics[width=6cm]{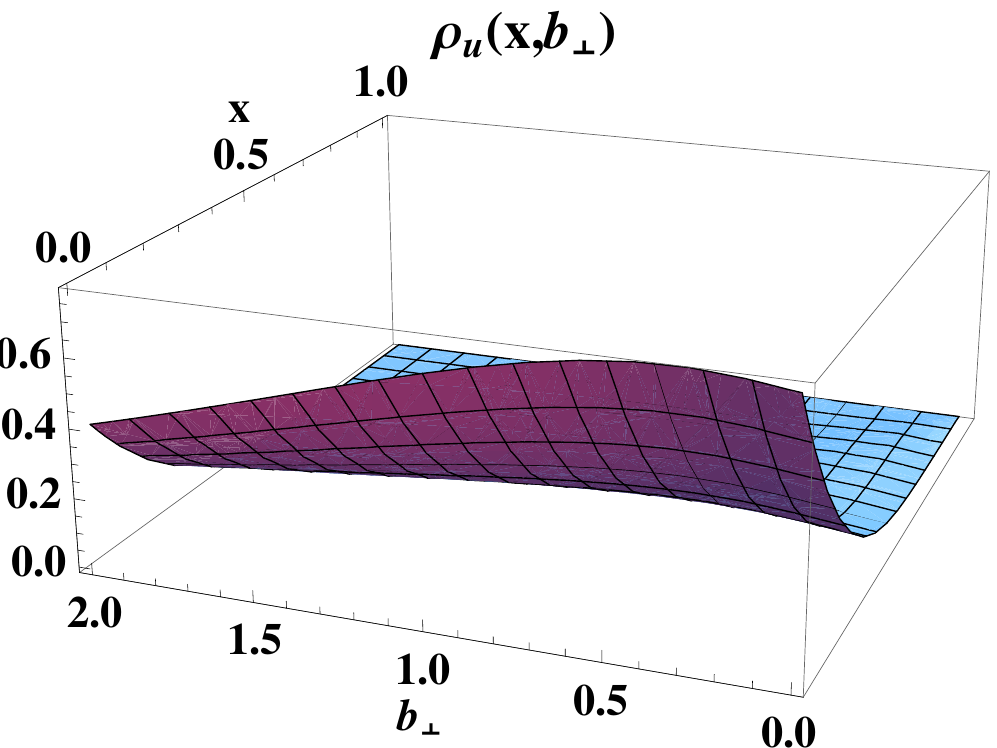}
\endminipage\hfill
\minipage{0.42\textwidth}
    \includegraphics[width=6cm]{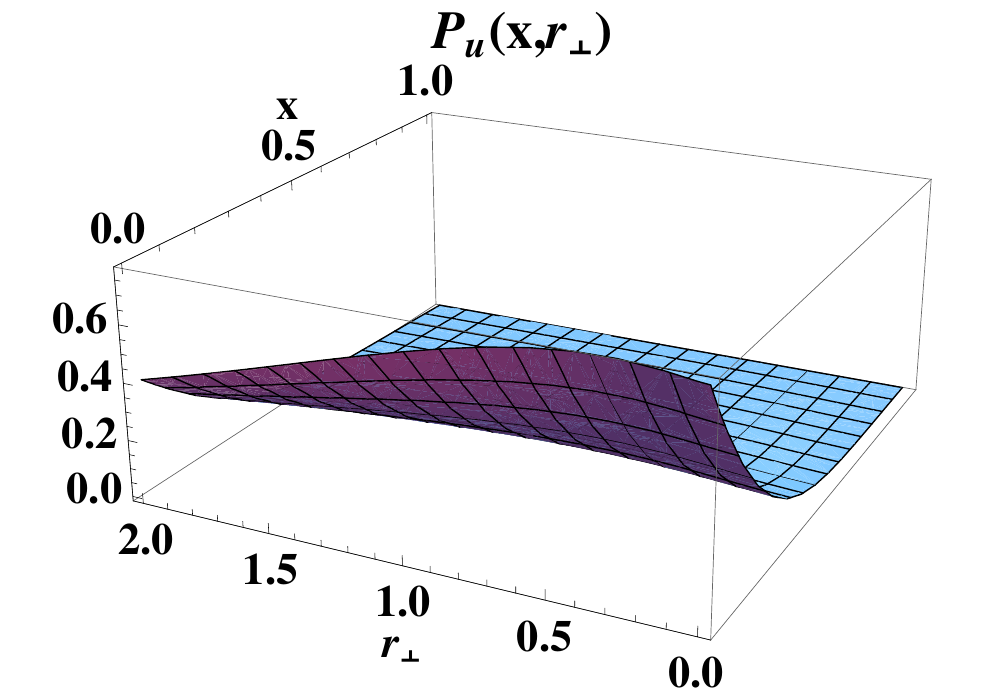}
  \endminipage\hfill
  \minipage{0.42\textwidth}
    \includegraphics[width=6cm]{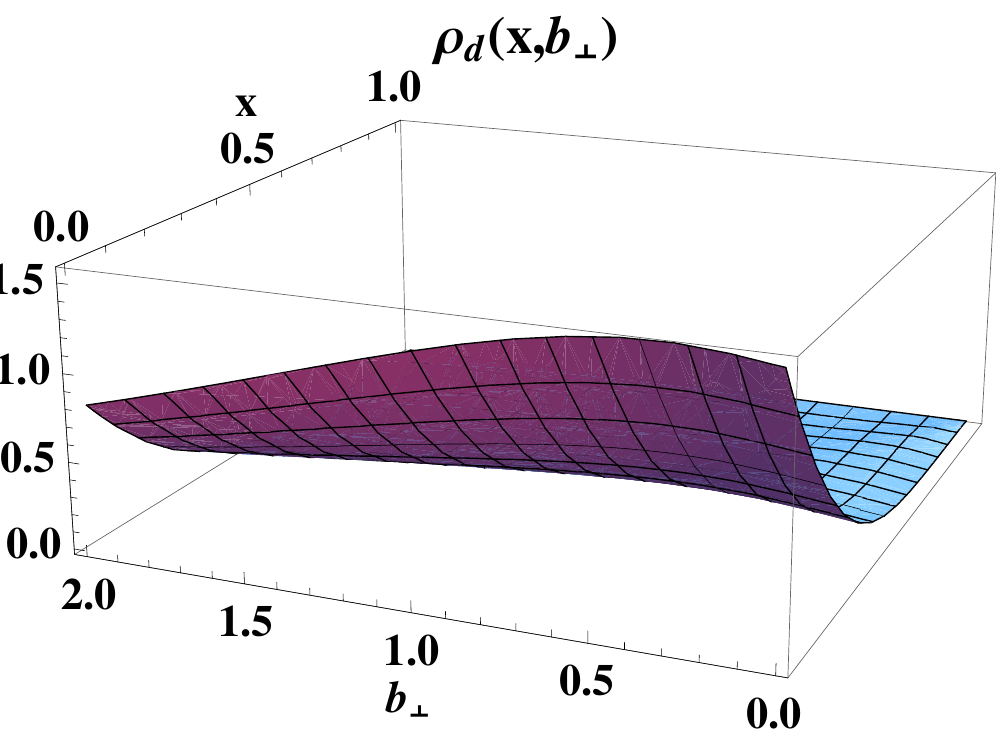}
  \endminipage\hfill
  \minipage{0.42\textwidth}
  \includegraphics[width=6cm]{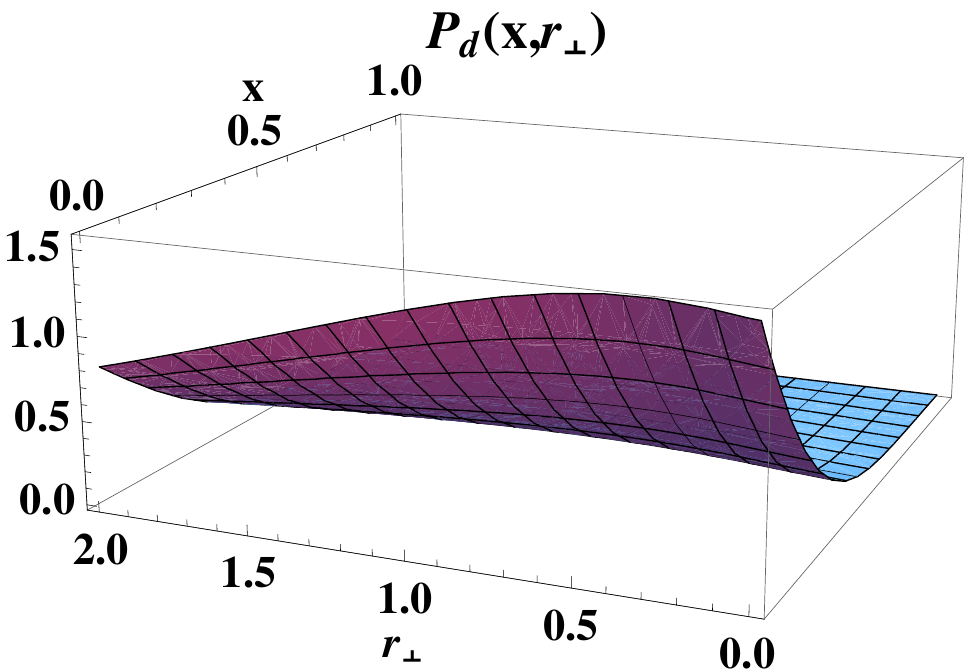}
\endminipage\hfill
  \caption{Results for $\rho_{u(d)}(x,b_\perp)$ (in $fm^{-2}$) and $P_{u(d)}(x,r_\perp)$ (in $fm^{-2}$) for the up (down) quark in  $b_\perp$ (in $fm$) and $r_\perp$ (in $fm$) space respectively. }
  \label{p_and_rho_up_down}
\end{figure}

\begin{figure}
\minipage{0.42\textwidth}
  \includegraphics[width=6cm]{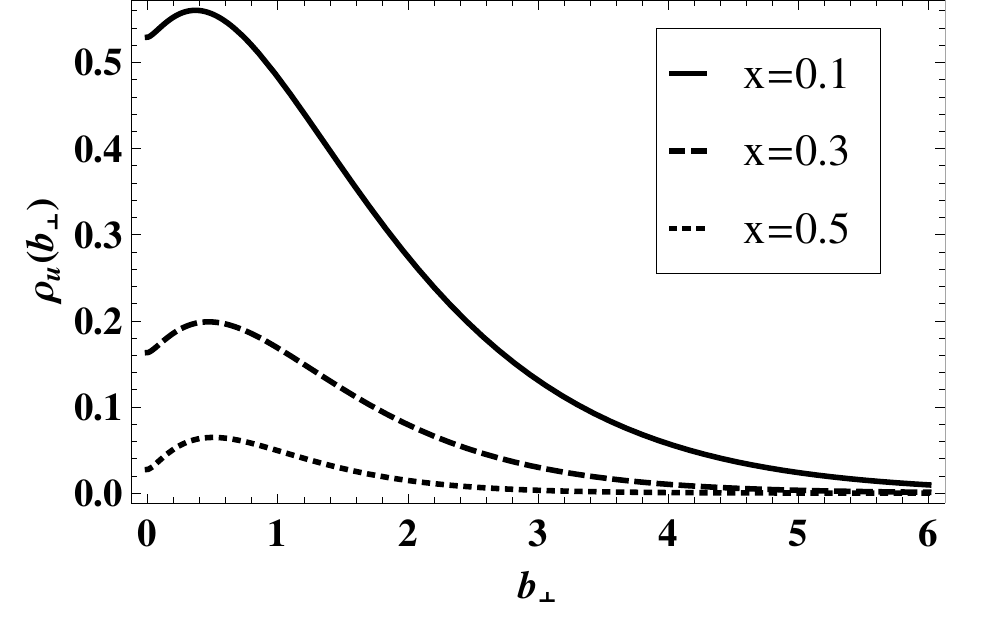}
\endminipage\hfill
\minipage{0.42\textwidth}
    \includegraphics[width=6cm]{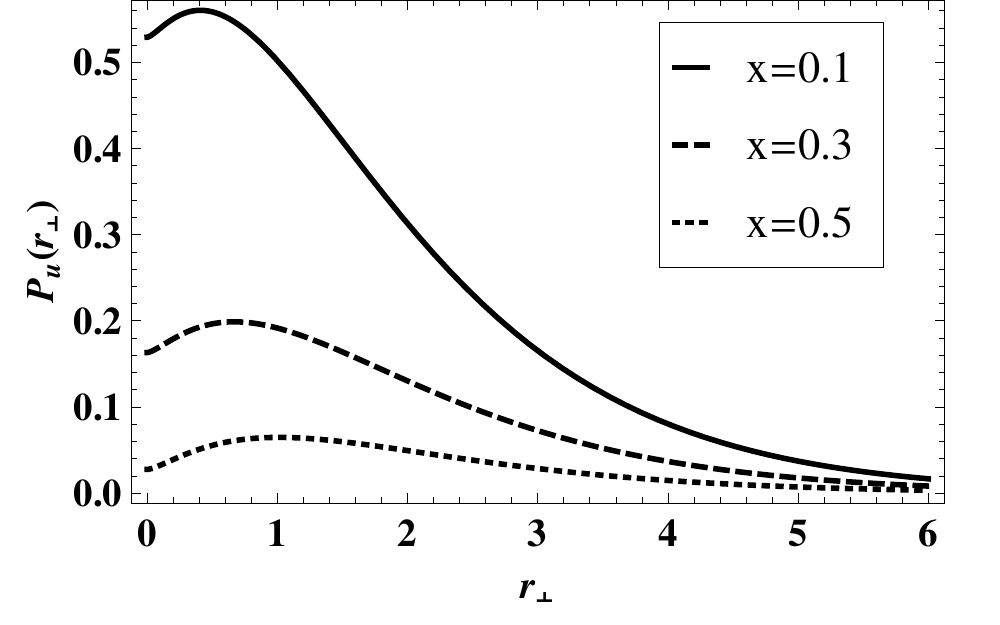}
  \endminipage\hfill
\minipage{0.42\textwidth}
  \includegraphics[width=6cm]{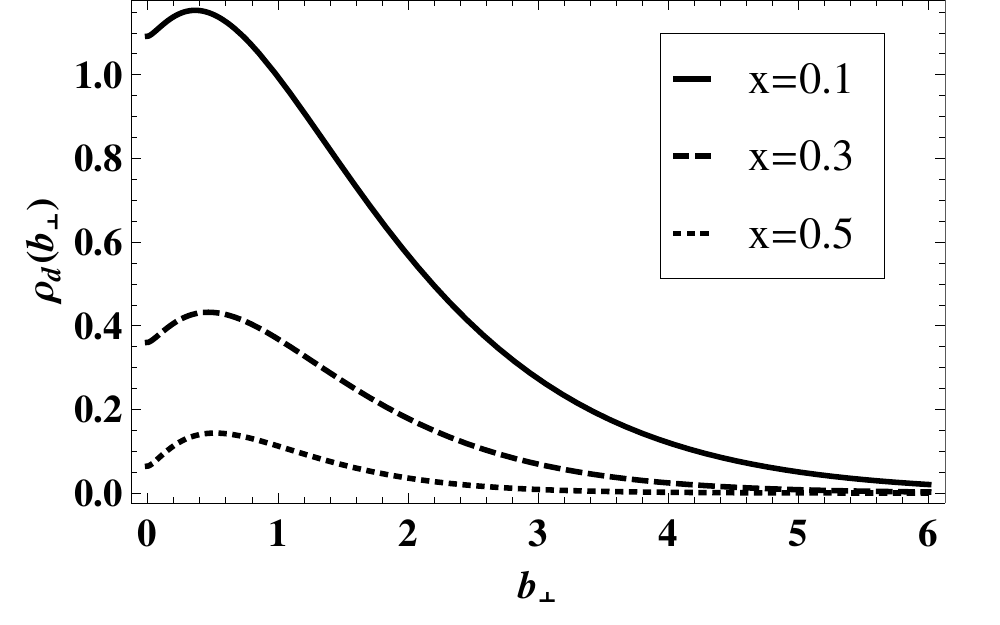}
\endminipage\hfill
\minipage{0.42\textwidth}
    \includegraphics[width=6cm]{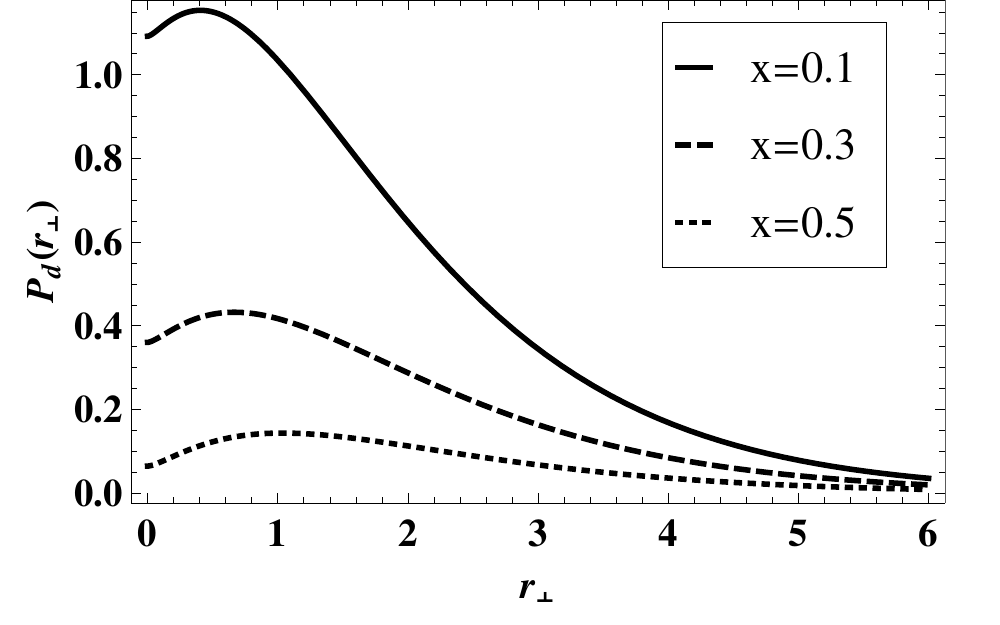}
  \endminipage\hfill
\caption{Results for $\rho_{u(d)}(x,b_\perp)$ (in $fm^{-2}$) and $P_{u(d)}(x,r_\perp)$ (in $fm^{-2}$) at $x=0.1$, $0.3$ and $0.5$. The units of $b_\perp$ and $r_\perp$ are in $fm$.}
 \label{p_and_rho_up_down_x}
\end{figure}

In Fig. \ref{p_and_rho_up_down}, we have presented the result for density distribution and charge distribution for {\it u} and {\it d} quarks respectively.  These plots give the complete spatial information about the nucleon. It is clear from the plots that charge  and density distributions increase with the increasing value of the impact parameter $b_\perp$ and the transverse coordinate $r_\perp$ respectively, reach a maxima and then start decreasing. The peak for the  charge distribution in impact parameter space is around $b_\perp\sim0.4 fm$ whereas the peak for the density distribution in transverse coordinate space is around $r_\perp\sim0.6 fm$ for both the $u$ and $d$ quarks. The magnitudes of $\rho_u(b_\perp)$ and $P_u(r_\perp)$ are however different from $\rho_d(b_\perp)$ and $P_d(r_\perp)$. A closer look at the figure reveals that the density distribution in transverse coordinate space falls off slowly as compared to the charge distribution in impact parameter space. This is due to the factor $\frac{b_\perp}{-1+x}$ appearing in Eq. (\ref{relation}).
To get a more clear picture of the distribution, we can plot the charge and density distribution at different values $x$. We have presented the results in Fig. \ref{p_and_rho_up_down_x} and as expected, the peak for different $x$ values occur the same value of $b_\perp$ and  $r_\perp$ but the magnitude of the peak decreases with the increasing value of $x$. It is important to mention here that since $x$ is the momentum fraction of the active quark, at $x=1$, the active quark carries all the momentum and the contribution from other partons is expected to be zero at this limit. The phenomenological analysis of existing data to determine the charge density of the proton and neutron has been discussed in Ref. \cite{chargedensity_p_n} where a model independent analysis of the charge density of partons has been done in the transverse plane for the nucleon. The charge density of the $d$ quarks is found to be larger than that of the $u$ quarks by 30\%. We also observe a large value of the $d$ quark density as compared to the $u$ quark density in both the spaces.

The difference between the distribution in the impact parameter and transverse coordinate space can be explained as follows. The GPDs $H$ and $E$ have dependence on $x$, $\xi$ and $t$ with $x$ being the longitudinal momentum fraction, $\xi$ the longitudinal momentum transfer and $t$ the invariant momentum transfer. We have fixed the longitudinal momentum transfer as $\xi=0$ in the present calculations which in result fixes the longitudinal momentum fraction $x$ in the impact parameter space $b_\perp$. Thus, one can get the exact information about the charge distribution in impact parameter space \cite{mukherjee}. However, in the case of transverse coordinate space, $r_\perp$ is a simple coordinate parameter directly obtained by taking Fourier transformation w.r.t. momentum $k_\perp$ and there is no parameter like $\xi$. Since the longitudinal momentum fraction $x$ cannot be fixed in the transverse coordinate space, one can get some contribution from longitudinal momentum as well and this basically leads to the slight difference of charge distribution  when compared with the distribution in the impact parameter  space. 

\section{Magnetization Density}
Another important quantity which has a form similar to the transverse charge density of a polarized nucleon is the nucleon magnetization density. Following Ref. \cite{gamiller,magmoment}, we can define  the anomalous magnetic moment $\mu_a$ as
\bea
\mu_a=\langle Y| \int d^2 b \ b_y \ q^{\dagger}_+ q_+ \ | Y \rangle ,
\label{ano_mag_moment1}
\eea
where $q$ is a quark field operator and $q_+=\gamma^0 \gamma^+ q$. The state for the transversely polarized target $Y$ is
\be
|Y \rangle=\frac{1}{\sqrt{2}}[|p^+, \vec{R}=\vec{0},+ \rangle + |p^+, \vec{R}=\vec{0},- \rangle],
\label{ano_mag_moment}
\ee
with $|p^+, \vec{R}=\vec{0},+ \rangle$ representing a transversely localized state.

Using the impact parameter distribution \cite{gamiller,impact3}, the magnetization density can be expressed as
\be
\rho_M(b_\perp)=\int \frac{d^2 q}{(2 \pi)^2}F_2(q^2) e^{-i \vec{q_\perp}. \vec{b_\perp}},
\label{rho_mag_moment}
\ee
which leads to the the anomalous magnetic moment
\be
\mu_a=\frac{1}{2 M} \int d^2b_\perp \ \rho_M(b_\perp).
\ee
Using Eq.  (\ref{ano_mag_moment1}) we can also express $\mu_a$ in terms of the anomalous  magnetization density $\tilde{\rho}_{M}(b_\perp)=- b_y \frac{\partial \rho_{M}(b_\perp)}{\partial b_y}$ as
\be
\mu_a=\frac{-1}{2 M} \int d^2 b \ b_y \frac{\partial \rho_{M}(b_\perp)}{\partial b_y}.
\label{ano_mag_moment2}
\ee
Substituting the value of $F_2(q^2)$, we can write
\bea
\tilde{\rho}_{M}(b_\perp)&=& \sin^2(\phi) b_\perp \int_{0}^{\infty} \frac{q^2 dq}{2 \pi} J_1(q \ b_\perp) F_2(q^2)\nonumber \\
&=& \sin^2(\phi) \ b_\perp \int_{0}^{\infty} \frac{q^2 dq}{2} J_1(q \ b_\perp)\Big(- 4 M^5 \int dx (M x-m) x (1-x)^3 I_2 \Big),
\label{rhomag}
\eea
with $\phi$ represents the angle between the direction of nucleon polarization (or the transverse magnetic field) and $b_\perp$.

\begin{figure}
    \includegraphics[width=8cm]{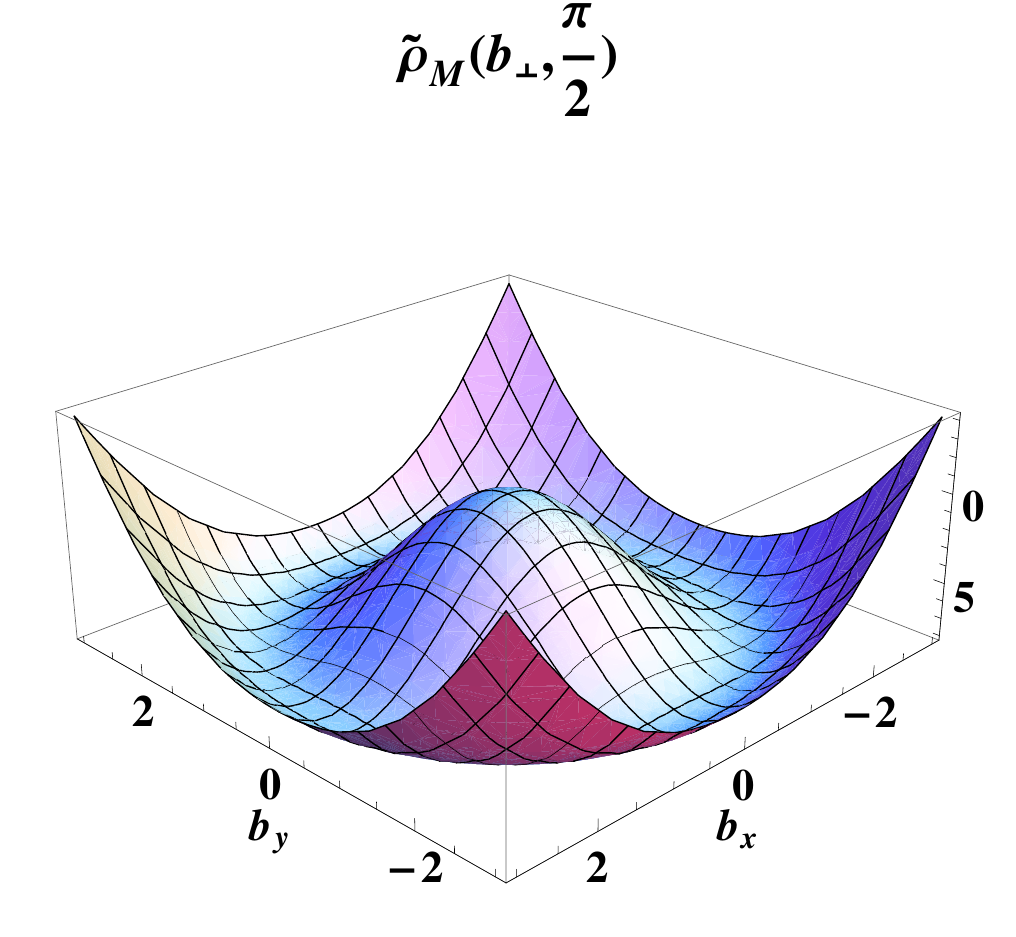}
 \caption{Plot for magnetization density $\tilde{\rho}_M(b_\perp)$ (in $fm^{-2}$) as a function of $b_\perp$ (in $fm$).}
 \label{mag3d}
 \end{figure}

In Fig. \ref{mag3d} we present the result for anomalous magnetization density as a function of $b_\perp$. Since the magnetization density also depends on $\phi$, we have taken $\phi=\frac{\pi}{2}$ in the present calculations where $\tilde{\rho}_{M}(b_\perp)$ will have the largest value. The anomalous magnetization density vanishes for $\phi=0$.  The anomalous magnetization density also vanishes for $b_\perp=0$ and shows a maxima around $b_\perp=2.5 fm$. For higher values of $b_\perp$ it falls off very rapidly leading to negative $\tilde{\rho}_{M}(b_\perp)$ values.

\section{charge density in AdS space}
In this section we have studied the charge density in AdS space. The LFWFs encode all the properties of hadron like bound state quark and gluon properties. A string amplitude $\Phi(z)$ can be defined on the fifth dimension in  AdS$_5$ space which can easily map the LFWFs of the hadrons \cite{adsspace,adsspace1,adsspace2,adsspace3}. It has been shown that a relation exists between the fifth dimension holographic variable $z$ and impact variable $\zeta$ which represents the measure of the transverse distance between the constituent within the hadrons. This relation provides an exact mapping at all energy scales between string modes in AdS and boundary states with a well defined number of partons. The string mode $\Phi(z)$ can be taken as probability amplitude for n-partons  at $\zeta=z$. The form factor in AdS is the overlap of the normalizable and the non-renormalizable modes \cite{adsspace}and can be expressed as
\be
F(q^2)=2 \pi \int_{0}^{1} dx \frac{(1-x)}{x} \int \zeta d\zeta J_0\left(\zeta q \sqrt{\frac{1-x}{x}}\right) \tilde{\rho}(x,\zeta).
\ee
The impact parameter space, obtained by taking the two dimensional Fourier conjugate with respect to  transverse momentum transfer, gives us the information about the  transverse structure of hadrons. On the other hand, due to correspondence between the fifth dimensional variable $z$ and the impact variable $\zeta$ for each n-parton Fock state in AdS space, it becomes very interesting to study the charge density in such a holographic model which explains the hadron spectrum remarkably. Since our aim is to study the charge density,  the impact parameters in both the spaces can be related to each other by the relation $\zeta^2=x(1-x) b_\perp^2$ for two partons. In the light of present work, the string modes can be further related to the QCD transverse charge density $\tilde{\rho}(x,\zeta)$ as follows
\be
\tilde{\rho}(x,\zeta)=\frac{R^3}{2 \pi}\frac{x}{1-x} e^{3 A(\zeta)} \frac{|\Phi(\zeta)|^2}{\zeta^4}.
\ee
Since the variable $\zeta$ represent the transverse distance between the constituent and holographic variable $z$ in AdS, we can take $\zeta=z$ leading to
\be
\tilde{\rho}_{n=2}(x,\zeta)=\frac{|\tilde{\psi}(x,\zeta)|^2}{(1-x)^2},
\ee
where
\be
\tilde{\psi}_{L,k}(x,\zeta)=B_{L,k} \sqrt{x(1-x)}J_L(\zeta \beta_{L,k} \Lambda_{QCD})\theta\Big(z\le \frac{1}{\Lambda_{QCD}}\Big),
\ee
and
\be
B_{L,k}=\Lambda_{QCD} \Big[(-1)^L \pi J_{1+L}(\beta_{L,k}) J_{1-L}(\beta_{L,k})\Big]^{-\frac{1}{2}}.
\ee
For two partons we can write the AdS density for ground state $(L=0,k=1)$ and excited state $(L=1,k=1)$ as follows
\bea
\tilde{\rho}_{n=2}(x,\zeta)(L=0,k=1)&=&\frac{x}{1-x}\frac{\Lambda_{QCD}^2 (J_0(\zeta \beta_{0,1} \Lambda_{QCD}))^2}{\pi (J_1(\beta_{(0,1)}))^2} , \nonumber\\
\tilde{\rho}_{n=2}(x,\zeta)(L=1,k=1)&=&- \frac{x}{1-x}\frac{\Lambda_{QCD}^2 (J_1(\zeta \beta_{1,1} \Lambda_{QCD}))^2}{\pi J_2(\beta_{(1,1)})J_0(\beta_{(1,1)})}.
\eea
\begin{figure}
\minipage{0.42\textwidth}
    \includegraphics[width=6cm]{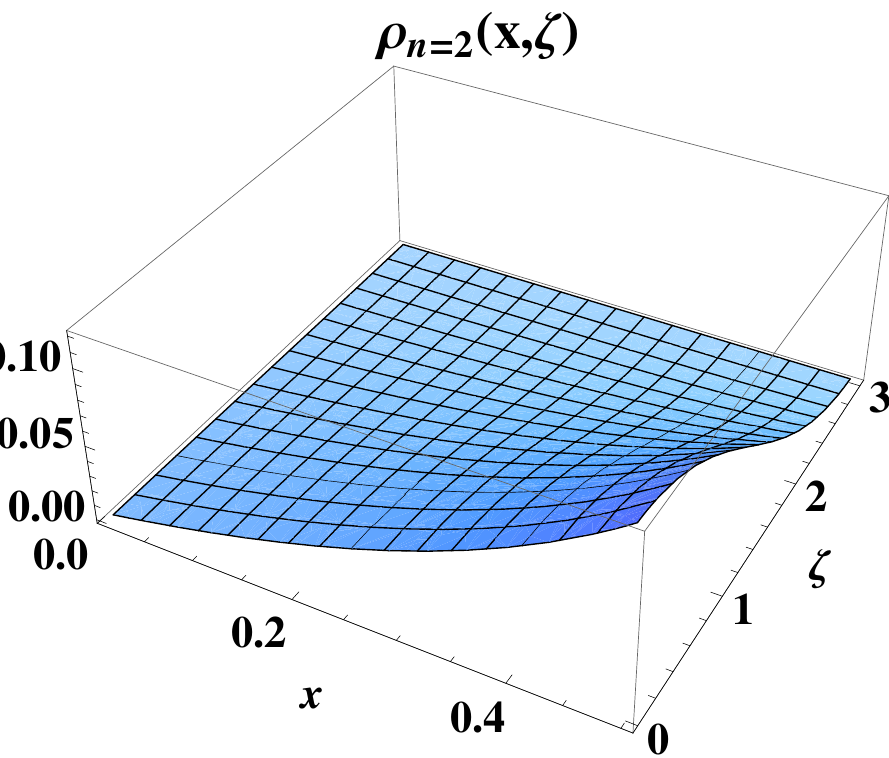}
  \endminipage\hfill
  \minipage{0.42\textwidth}
  \includegraphics[width=6cm]{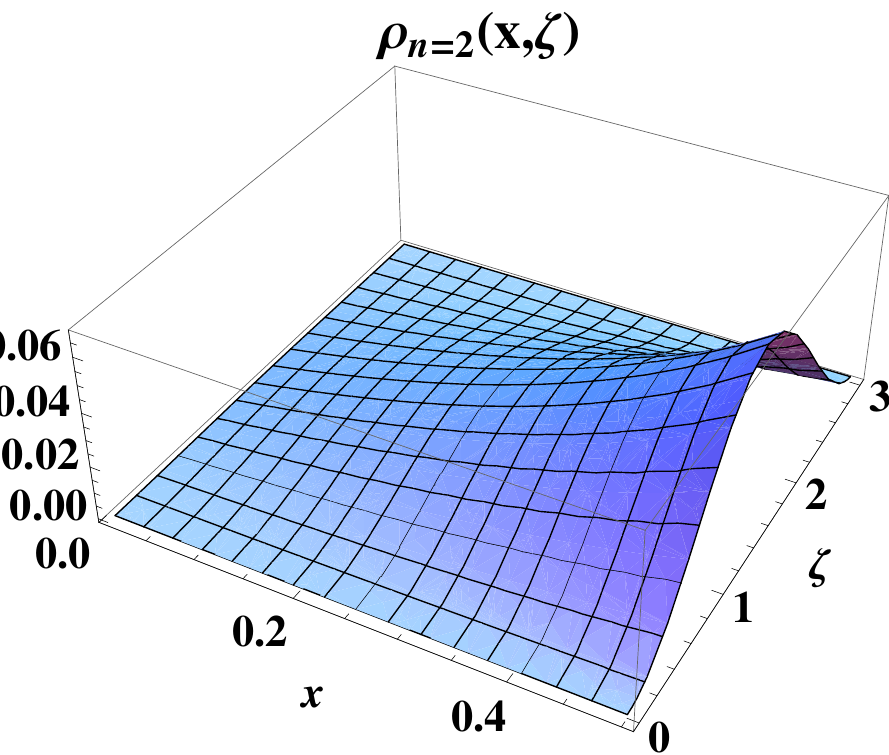}
\endminipage\hfill
\caption{Results for AdS density as a function of $x$ and $\zeta$ (in GeV$^{-1}$) with $\Lambda_{QCD}=0.32$GeV for ground state $L=0,k=1$ (LHS) and orbital excited state $L=1,k=1$ (RHS). }
 \label{AdS_density}
\end{figure}
In Fig. \ref{AdS_density} we have plotted the AdS density with $\Lambda_{QCD}=0.32$GeV and $z \le \frac{1}{\Lambda_{QCD}}$. We observe that for ground state $(L=0,k=1)$ the magnitude of charge density decreases as the value of $\zeta$ increases which is primarily due to the presence of the Bessel function in the LFWFs. The magnitude is maximum when $\zeta=0$ which implies that when the transverse distance between the constituent within the hadrons is minimum the density is maximum. In the first excited state  $(L=1,k=1)$, we have obtained a peak for charge density in AdS space for a comparatively higher value of $\zeta$. The density is also dependent on the variable $x$ and it is observed that for the lower values of $x$, the density is zero. The peak occurs only at higher values of $x$. More secondary peaks will be observed for the charge density as the value of $L$ and $k$ increase further on this scale.

\section{conclusions}

In the present work, we have considered the light-cone Fock state representation of the composite system consisting of external and internal fermion state with respective masses $M$ and $m$. The mass of the vector constituent is taken to be $\lambda$. The electromagnetic Dirac and Pauli form factors as well as the Sachs form factors $(G_E(Q^2))$ and $(G_M(Q^2))$ have been calculated from the overlap of LFWFs.  The nucleon charge density has been obtained for the case of unpolarized nucleon in transverse coordinate space as well as in the impact parameter space. Further, a relation has been developed between the charge distribution of the quarks inside nucleon in the transverse coordinate space as well as in the impact parameter space particularly for the $u$ and $d$ quark distribution by using a fitted set of parameters. It is found that density distribution in transverse coordinate space falls off slowly as compared to the charge distribution in impact parameter space but the magnitude of the peak decreases with the increasing value of the momentum fraction of the active quark.  This is primarily because of some contribution of the longitudinal momentum fraction $x$ in  the transverse coordinate space whereas in the impact parameter space, the longitudinal momentum fraction $x$ can be fixed. Further as $x\rightarrow1$, the contribution from other partons is expected to be zero and the active quark  carries all the momentum. This investigation gives a deeper understanding of both the spatial and transverse distribution of the quark densities inside the nucleon. We have also studied the anomalous magnetization density of the nucleon obtained from the spin-flip matrix element of the $J^+$ current. The anomalous magnetization density vanishes when the angle between the direction of  nucleon polarization and $b_\perp$ is zero as well as when $b_\perp=0$. It maxima occurs around $b_\perp=2.5 fm$ and for higher values $\tilde{\rho}_{M}(b_\perp)$ falls off very rapidly. Further, the calculations were extended to predict the results of the QCD transverse AdS charge density inspired from the holographic QCD model and it is observed that for the ground state when the transverse distance between the constituent within the hadrons is minimum the density is maximum. For the first excited state, a peak for charge density is observed at a comparatively higher value of transverse distance. Even more secondary peaks will be observed for higher excited states.

\section*{ACKNOWLEDGMENTS}
Authors acknowledge helpful discussions with S.J. Brodsky. HD would like to thank Department of Science and Technology, Government of India for financial support.
\appendix
\section{}
If we have any two arbitrary functions $\psi_c(\vec{k}_\perp)$ and $\psi_d(\vec{k}_\perp)$, their Fourier transforms can be given as
\be
\psi_c(\vec{k}_\perp)= \int d^2 \vec{r}_\perp e^{- i \vec{k}_\perp \cdot \vec{r}_\perp} \tilde{\psi_c}(\vec{r}_\perp), \
\psi_d(\vec{k}_\perp)= \int d^2 \vec{r}_\perp e^{- i \vec{k}_\perp \cdot \vec{r}_\perp} \tilde{\psi_d}(\vec{r}_\perp).
\ee
The form factor can then be expressed as
\be
F(\vec{q}_\perp)= \int \frac{d^2\vec{k}_\perp}{(2\pi)^2} \psi_c^*(\vec{k}_\perp - \vec{q}_\perp) \psi_d(\vec{k}_\perp),
\ee
which, in Fourier conjugate space, becomes diagonal
\be
\int \frac{d^2\vec{q}_\perp}{(2\pi)^2}e^{i \vec{q}_\perp \cdot \vec{r}_\perp} F(\vec{q}_\perp)= \tilde{\psi_c}^* (\vec{r}_\perp) \tilde{\psi_d} (\vec{r}_\perp).
\ee
Similarly, for
\be
G(\vec{q}_\perp)= \int \frac{d^2\vec{k}_\perp}{(2\pi)^2} \psi^*(\vec{k}_\perp - y \ \vec{q}_\perp) \psi(\vec{k}_\perp),
\ee
we have
\be
\int \frac{d^2\vec{q}_\perp}{(2\pi)^2} e^{i \vec{q}_\perp \cdot \vec{b}_\perp} G(\vec{q}_\perp)=\frac{1}{|y|^2} \tilde{\psi}^*\left(\frac{\vec{b}_\perp}{y}\right) \tilde{\psi}\left(\frac{\vec{b}_\perp}{y}\right).
\ee

\end{document}